\documentclass[graybox]{svmult}

\usepackage{mathptmx}       
\usepackage{helvet}         
\usepackage{courier}        
\usepackage{makeidx}         
\usepackage{graphicx}        
\usepackage{multicol}        

\makeindex             

\begin{document}

\title*{Variable stars in the globular cluster M\,79}
\author{Grzegorz Kopacki and Andrzej Pigulski}
\institute{Grzegorz Kopacki and Andrzej Pigulski 
 \at Instytut Astronomiczny Uniwersytetu Wroc\l{}awskiego, Kopernika 11, 51-622 Wroc\l{}aw, Poland,
 \email{kopacki@astro.uni.wroc.pl}
}
%
%
\maketitle

\abstract*{We present preliminary results obtained from analysis of the $VI$ photometry of the 
globular cluster M\,79. Stellar variability survey 
performed with the image subtraction method yielded six new pulsating 
stars: two of RR Lyrae type, three SX Phoenicis stars and even one W 
Virginis star. Using all eleven RR Lyrae stars known in the cluster we find 
that M\,79 is Oosterhoff type II globular cluster.}

\abstract{We present preliminary results obtained from analysis of the $VI$ photometry of the 
globular cluster M\,79. Stellar variability survey 
performed with the image subtraction method yielded six new pulsating 
stars: two of RR Lyrae type, three SX Phoenicis stars and even one W 
Virginis star. Using all eleven RR Lyrae stars known in the cluster we find 
that M\,79 is Oosterhoff type II globular cluster.}

\section{Introduction, Observations and Results}

 In the last two decades we observed a rapid increase in the number of 
 variable stars detected in Galactic globular clusters. The main 
 reason for this was the invention of the image subtraction method (ISM, \cite{ala98}) and its application to CCD data obtained 
 with small telescopes (e.g.\ \cite{kop09}). 
 The ISM enables making a complete inventory of
 bright variable stars, such as RR Lyrae variables, because it works
 well in crowded stellar fields like the cluster core. However, there are
 still many globular clusters  poorly searched for variable stars, 
 especially pulsating stars of the RR Lyrae and SX Phoenicis types.

 Here we present results of a variability analysis for the 
 southern globular cluster M\,79 (NGC\,1904) of intermediate metallicity
 ([Fe/H]${}=-1.57$). The Catalogue of Variable
 Stars in Globular Clusters (CVSGC, \cite{cle01}) listed
 13 objects in the field of this cluster including nine RR Lyrae stars,
 one semiregular variable and three stars suspected for variability.

 We used two sets of observations. The first one consisted of 690
 $V$-filter and 230 $I_{\rm C}$-filter CCD frames obtained during 
 one-month observing run in Feb/Apr, 2008 using 40-inch telescope 
 at SSO, Australia.
 The other one included 80 $V$-filter CCD frames acquired during one 
 week of observation in Apr, 2008 using 40-inch telescope at SAAO, 
 South Africa.

 We have detected two new RR Lyrae stars, both of the RRc type.
 The mean period of RRab stars in M\,79 is equal to $\langle P\rangle_{\rm ab}=0.69$ d, and
 relative percentage of RRc stars amounts to $N_{\rm c}/(N_{\rm ab}+N_{\rm c})=44$ \%.
 With these values we conclude that M\,79 belongs to the Oosterhoff's II group of globular clusters.
 
 We show that v7 is a W Virginis-type star. Moreover, three SX Phoenicis stars were found among 
 cluster's blue stragglers. One of them, n18, and RRc star v9 are multiperiodic pulsators.
 The period ratio for n18 indicates that this SX Phoenicis star is a double-mode radial pulsator.
 Positions and periods of all observed periodic stars are given in Table~\ref{tab:1}.
 Irregular light variations were also found in a dozen of red giants located
 at the tip of the cluster's red giant branch.

\begin{table}
\caption{Equatorial coordinates and periods for periodic stars in M\,79}
\label{tab:1}       
\begin{tabular}{p{0.8cm}p{2cm}p{2cm}p{2cm}p{0.8cm}}
\hline\noalign{\smallskip}
 Var& $\alpha_{2000}$ [$^{\rm h}$ $^{\rm m}$ $^{\rm s}$]& $\delta_{2000}$ [$^\circ$ $^\prime$ $^{\prime\prime}$]& P [d]& Type\\
\noalign{\smallskip}\svhline\noalign{\smallskip}
  v3& 5 24 13.54& $-$24 32 29.1&  0.73602&   RRab\\ 
  v4& 5 24 17.77& $-$24 32 16.2&  0.63339&   RRab\\ 
  v5& 5 24 10.23& $-$24 31 03.6&  0.66894&   RRab\\ 
  v6& 5 24 06.03& $-$24 29 32.9&  0.339065&  RRc\\ 
  v7& 5 24 12.68& $-$24 31 41.9& 13.946&     W\,Vir\\ 
  v9& 5 24 12.58& $-$24 31 52.6&  0.37905&   RRc\\
    &           &            &  0.36099&\\
    &           &            &  0.37049&\\
 v10& 5 24 12.13& $-$24 31 34.5&  0.72894&   RRab\\
 v11& 5 24 11.93& $-$24 31 34.6&  0.8232&    RRab\\ 
 v12& 5 24 11.35& $-$24 31 28.3&  0.32394&   RRc\\ 
 v13& 5 24 10.59& $-$24 31 11.5&  0.68931&   RRab\\ 
 n14& 5 23 23.74& $-$24 27 46.3&  0.309058&  RRc\\ 
 n15& 5 24 07.77& $-$24 31 00.3&  0.323758&  RRc\\ 
 n16& 5 24 09.97& $-$24 31 07.3&  0.038763&  SX\,Phe\\ 
 n17& 5 24 14.15& $-$24 33 20.7&  0.044803&  SX\,Phe\\ 
 n18& 5 24 10.86& $-$24 31 11.8&  0.050276&  SX\,Phe\\ 
    &           &            &  0.039169&\\
\noalign{\smallskip}\hline\noalign{\smallskip}
\end{tabular}
\end{table}

\begin{acknowledgement}
This work was supported by Polish Ministry of Science grant N203 014 31/2650.
\end{acknowledgement}

\end{document}